\documentstyle[11pt]{article}

\newcommand{\be}{\begin{equation}}
\newcommand{\ee}{\end{equation}}
\newcommand{\bea}{\begin{eqnarray*}}
\newcommand{\eea}{\end{eqnarray*}}
\newcommand{\ba}{\begin{eqnarray}}
\newcommand{\ea}{\end{eqnarray}}
\renewcommand{\baselinestretch}{2}
\begin{document}
\begin{flushright} UCL-IPT-98-04
\end{flushright}

\vspace*{5mm}

\begin{center}
\Large{\bf Hadronic Phases and Isospin Amplitudes in $ D (B) \to \pi \pi $
and  $ D
 (B) \to K
\bar K $ Decays}
\end{center}

\vspace*{15mm}

\begin{center}
\Large{\it J.-M. G\'erard, J. Pestieau and J. Weyers }
\end{center}

\vspace*{5mm}

\begin{center}
Institut de Physique Th\'eorique\\
Universit\'e catholique de Louvain \\
 B-1348 Louvain-la-Neuve,
Belgium
\end{center}

\thispagestyle{empty}

\vspace*{40mm}

\renewcommand{\baselinestretch}{1.4}

\begin{abstract}

Hadronic phases in $\pi \pi$ and $K \bar K$ channels are calculated \`a la
Regge.  At the $D$ mass
one finds $\delta _{\pi\pi} \simeq \frac{\pi}{3} $ and $\delta_{K \bar K}
\simeq -
\frac{\pi}{6}$ in  good agreement with the CLEO data while at the $B$ mass these
angles are predicted to be, respectively,
11$^\circ$ and -7$^\circ$. With the hadronic phase $e^{i\delta_{ K \bar
K}}$ taken
into account, a quark diagram decomposition of the isospin invariant
amplitudes in $D
\to K
\bar K$ decays  fits the data provided the  exchange
  diagram contribution is about $1/3$ of the tree level one.
\end{abstract}

\renewcommand{\baselinestretch}{2}

\newpage

\section{Introduction}

To extract information on weak interaction parameters from non leptonic two body
decays of the $D$ and $B$ mesons, it is crucial to understand the final hadronic
effects which are at work in these decays.  For ($K \pi$), ($\pi \pi$) and
($K \bar
K$) decay modes, the important hadronic parameter is an angle $\delta$
which is the
difference between $s$-wave phase shifts in the appropriate isospin invariant
amplitudes.

In a previous paper \cite{1} we used a Regge model to determine $\delta_{K
\pi}$ as a
function of energy.  Good agreement with data at the $D$ mass \cite{2} was
obtained
and
$\delta_{K
\pi}$ is predicted to be around $20^\circ$ at the $B$ mass.

In this letter we extend
this Regge analysis to ($\pi \pi$) and ($K \bar K$) channels and determine
$\delta_{\pi \pi}(s)$ and $\delta_{K \bar K}(s)$.  At the $D$ mass, we find
$\delta_{\pi \pi}(m^2_D) \simeq \frac{\pi}{3}$ and $\delta_{K \bar K}
(m^2_D) \simeq - \frac{\pi}{6}$ once again in agreement with the
data\cite{3}\cite{2}.  At the
$B$ mass these angles are predicted to be of the order of $11^\circ$ and
$-7^\circ$ respectively, implying that hadronic effects in $B$ decays remain
important.  Details of the derivation of these results are given in Section 2.
	
With hadronic phases thus determined it becomes interesting to compare a
quark diagram
decomposition of isospin invariant amplitudes with the data.  In Section 3,
we argue
that for $D
\to K \bar K$ decays a fit to the data \cite{2} implies that the contribution of
exchange quark diagrams be of the order of 1/3 of the tree-level one.

\section{$\delta_{\pi \pi}(s)$ and $\delta_{K \bar K} (s)$ in a Regge Model}

In $\pi \pi \to \pi \pi$ scattering, isospin eigenamplitudes ($I = 0,1,2$)
in the
$s,t,u$-channels are related by the crossing matrices
\be
\left(\begin{array}{c}
A^s_0 \\
A^s_1 \\
A^s_2
\end{array}
\right) =
\left(\begin{array}{clrr}
1/3 & \ \ \ 1 & 5/3 \\
1/3 & \ \ 1/2 & -5/6 \\
1/3 & -1/2 & 1/6
\end{array}
\right)
\left(\begin{array}{c}
A^t_0 \\
A^t_1 \\
A^t_2
\end{array}
\right) =
\left(\begin{array}{rlrrr}
1/3 & \ \ \ \  1 & 5/3 \\
-1/3 & -1/2 & 5/6 \\
1/3 & -1/2 & 1/6
\end{array}
\right)
\left(\begin{array}{c}
A^u_0 \\
A^u_1 \\
A^u_2
\end{array}
\right)
\ee
while for $K \bar K \to K \bar K$ scattering ($I = 0,1$) the $s-t$ crossing
matrix reads
\be
\left(\begin{array}{c}
\tilde A^s_0 \\
\tilde A^s_1
\end{array}
\right) =
\left(\begin{array}{rrrr}
1/2 & 3/2 \\
1/2 & -1/2
\end{array}
\right)
\left(\begin{array}{c}
\tilde A^t_0 \\
\tilde A^t_1
\end{array}
\right).
\ee

The basic physical idea of a Regge  model is that the high energy behaviour of
$s$-channel amplitudes is determined by "exchanges" in the crossed
channels.  For
$\pi \pi$ scattering, the dominant exchanges in the $t$-channel are the
Pomeron $(P)$
and the exchange degenerate $\rho-f_2$ trajectories while in $K \bar K$
scattering
one must also add the exchange degenerate $\omega - a_2$ trajectories.  The
$u$-channel exchanges in $\pi \pi$ scattering are identical to the
$t$-channel ones
$(P,\rho, f_2)$ while in  $K \bar K$ scattering  there are no exchanges in the
(exotic) $u$ channel ($K K \to K K$).
	
In the energy range (3 GeV$^2$
${<\hspace{-3.5mm}\raisebox{-1mm}{$\scriptstyle \sim$}}$
$ s$
${<\hspace{-3.5mm}\raisebox{-1mm}{$\scriptstyle \sim$}}$ 35 GeV$^2$) which is of
interest to us, the Pomeron ($P$) contribution to the isoscalar $t$-channel
amplitude is phenomenologically well described by the formula
\be
A_P = i \beta_P(0) e^{ib_pt} s
\ee
where the residue $\beta_P (0)$ and slope $b_P$ depend on the scattering
process considered.
The $\rho,f_2, \omega, a_2$ Regge trajectories are all degenerate  i.e.
\be
\alpha_\rho (t) = \alpha_{f_2} (t) = \alpha_\omega (t) = \alpha_{a_2} (t) =
\frac{1}{2} + t
\ee
The $\rho$ and $\omega$ trajectories have negative signatures while the
$f_2$ and $a_2$
trajectories are of positive signature.
	
The effective $\rho$ trajectory contribution to the isovector $t$-channel
amplitude is
written as
\be
A_\rho = \frac{\bar \beta \rho}{\sqrt{\pi}} (1 + ie^{-i \pi t}) s^{0.5 + t}
\ee
while the $a_2$ contribution to ${\widetilde A_1^t}$ reads
\be
A_{a_2} = \frac{\widetilde{\bar \beta} a_2}{\sqrt{\pi}} (-1 + i e^{-i \pi
t}) s^{0.5 +
t}.
\ee
Similar expressions are used for the effective $\omega$ and $f_2$ trajectory
contributions to the isoscalar $t$-channel amplitude.

The residues of these trajectories are related as follows :

\par\noindent
a) in $\pi \pi$ scattering
\be
\bar \beta_{f_2} = \frac{3}{2} {\bar \beta_\rho}
\ee

\par\noindent
b) in $K \bar K$ scattering
\ba
\widetilde{\bar \beta}_{f_2} &=& \widetilde{\bar \beta}_\rho
\\
\widetilde{\bar \beta}_{a_2} &=& \widetilde{\bar \beta}_\omega.
\ea

Furthermore, $SU(3)$ symmetry and ideal mixing give the additionnal relation
\be
\widetilde{\bar \beta}_\rho =  \widetilde{\bar \beta}_\omega.
\ee

Eqs (7)-(9) follow from ``duality" : the scattering processes $(\pi^+ \pi^+
\to \pi^+
\pi^+) (A_2^s)$ and ($KK \to KK$) are purely diffractive, hence the
imaginary part
of the Regge trajectory contributions to these processes must cancel \cite{4}.

Using Eqs (1) (3)-(5) and (7) our Regge model for $\pi \pi$ scattering near
the forward
direction ($t$ small) reads :
\ba
A_0^s (s \ {\rm large}, {\rm small} \  t) &=& \frac{i}{3} \beta_P(0)
e^{b_pt} s +
\frac{1}{2}
\frac{\bar
\beta_\rho}{\sqrt{\pi}} s^{0.5 + t} + \frac {3i}{2\sqrt{\pi}} \bar \beta_\rho
e^{-i\pi t} s^{0.5 +t}
\\
A_2^s (s \ {\rm large}, {\rm small} \  t) &=& \frac{i}{3} \beta_P(0)
e^{b_pt} s -
\frac{\bar \beta_\rho}{\sqrt{\pi}} s^{0.5 + t}.
\ea
In the backward direction ($u$ small) exactly the same formulae hold with $t$
replaced by $u$.

Similarly, for $K \bar K \to K \bar K$ scattering one obtains from Eq.(2),
using the
relations Eqs (8)-(10), that
\ba
\widetilde {A_0^s} (s \ {\rm large}, {\rm small} \  t) &=& \frac {i}{2} \widetilde
\beta_P (0) e^{\widetilde \beta_P t}s + \frac {4i \widetilde{\bar
\beta}_\rho}{\sqrt{\pi}} e^{-i\pi t}s^{0.5 + t}
\\
\widetilde {A_1^s} (s \ {\rm large}, {\rm small} \ t) &=& \frac {i}{2}
\widetilde
\beta_P (0) e^{\tilde \beta_P t}s.
\ea
From Eqs (11)-(14) we compute the $l=0$ partial wave amplitudes and find, up to
irrelevant overall real factors
\ba
a_0 (s) &=& \frac{i}{3} \frac {\beta_P(0)}{b_P} s
+ \frac{\bar \beta_\rho}{2\sqrt{\pi}} \frac{s^{1/2}}{\ln s}
+ \frac{3i}{2\sqrt{\pi}} \bar \beta_\rho \frac {(\ln s) + i \pi}{(\ln s)^2
+\pi^2}
s^{1/2}
\\
a_2 (s) &=& \frac{i}{3} \frac {\beta_P(0)}{b_P} s - \frac{\bar
\beta_\rho}{\sqrt{\pi}}
\frac{s^{1/2}}{\ln s}
\\
\widetilde a_0 (s) &=& \frac{i}{2} \frac {\widetilde \beta_P(0)}{\widetilde
b_P} s
+ \frac{4i \widetilde{\bar \beta_\rho}}{\sqrt{\pi}} \frac {(\ln s) + i
\pi}{(\ln s)^2
+ \pi^2} s^{1/2}
\\
\widetilde a_1 (s) &=& \frac{i}{2} \frac {\widetilde \beta_P(0)}{\widetilde
b_P} s.
\ea
The $u$-channel contributions in Eqs. (15)-(16) are identical to the
$t$-channel ones and we have dropped a (common) factor of 2 in these equations.
	
Clearly the phases $e^{i\delta_0}$ and $e^{i\delta_2}$ of $a_0 (s)$ and
$a_2 (s)$
depend on the phenomenological parameter
\be
x_{\pi \pi} = \frac {\sqrt{\pi}\beta_P (0)}{\bar \beta_\rho (0)}\frac {1}{b_P}
\ee
and similarly  $e^{i\widetilde \delta_0}$ and $e^{i\widetilde \delta_1}$
depend on
\be
x_{K \bar K} = \frac {\sqrt{\pi}\widetilde \beta_P (0)}{\widetilde{\bar
\beta}_\rho
(0)}\frac {1}{\tilde b_P}.
\ee

From the fits given in references \cite{5} and \cite{6}\cite{7} we extract
the values
\ba
x_{\pi \pi} &=& 0.69 \pm 0.10
\\
x_{K \bar K} &=&  1.72 \pm 0.30.
\ea
With these values we obtain respectively
\ba
\delta_{\pi \pi}(m_D^2) &=& \delta_2(m_D^2) - \delta_0(m_D^2) = 60^\circ
\pm 4^\circ
\\
\delta_{K \bar K}(m_D^2) &=& \tilde \delta_1 (m_D^2) -
\tilde \delta_0 (m_D^2) = -29^\circ \pm 4^\circ
\ea
and predict
\ba
\delta_{\pi \pi}(m_B^2) &=& 11^\circ \pm 2^\circ
\\
\delta_{K \bar K}(m_B^2) &=& -7^\circ \pm 1^\circ.
\ea
At the $D$ mass, the experimental values given by the CLEO collaboration
\cite{3}\cite{2} are,
\ba
\delta_{\pi \pi}(m_D^2) &=& 82^\circ \pm 10^\circ
\\
\delta_{K \bar K}(m_D^2) &=& \pm (24^\circ \pm 13^\circ).
\ea

Clearly our Regge model calculation of these phases is in good agreement
with the
data as announced previously.

In summary, for ($\pi \pi$) and ($K\bar K$) decay channels as well as for
($K \pi$)
channels, hadronic angles are correctly predicted at the $D$ mass by a
Regge model
and are found to be quite sizeable at the $B$ mass : hadronic effects
simply cannot
be ignored in $B$ decays.

\section{Isospin Amplitudes and Quark Diagrams in ($D \to K \bar K$) Decays}

Having determined the hadronic phase $\delta_{K \bar K}$, we now illustrate the
strategy advocated in Ref.\cite{8}  to analyze the $D \to K \bar K$ data.

In lowest order the weak Hamiltonian responsible for ($D \to K \bar K$) decays
contains an isodoublet ($H_W^{1/2}$) and an isoquadruplet ($H_W^{3/2}$)
part.  With
the reduced matrix elements
\ba
w_1 &=& \ll D \mid H_W^{1/2} \mid (K \bar K) I = 1 \gg
\\
w_0 &=& \ll D \mid H_W^{1/2} \mid (K \bar K) I = 0 \gg
\\
v_1 &=& \ll D \mid H_W^{3/2} \mid (K \bar K) I = 1 \gg
\ea
and the {\em hadronic} angle $\delta_{K \bar K} = \delta_1 - \delta_0$ one
readily
obtains, up to an overall phase factor
\ba
A(D^+ \to K^+ \bar K^0) &=& - \frac{v_1}{2} + w_1
\\
A(D^0 \to K^+  K^-) &=&  \frac{v_1}{2} + \frac{w_1}{2} + \frac{w_0}{2}e^{-i
\delta_{K
\bar K}}
\\
A(D^0 \to K^0 \bar K^0) &=&  - \frac{v_1}{2} - \frac{w_1}{2} +
\frac{w_0}{2}e^{-i
\delta_{K \bar K}}.
\ea

We assume again \cite{8} all reduced matrix elements to be real and
expressed in
terms of {\em quark} diagrams classified following their $(1/N$-inspired)
topology.

In this ``phenomenological" picture where we keep the explicit $(V-A)$
times $(V-A)$ $ W^\pm$
propagations, the annihilation diagram $(A)$ is helicity-suppressed and the
$b,s$ and
$d$ quarks are exchanged in the penguin diagrams $(P_q$). On the contrary, in the
``formal" language \cite{9} the effects of $W^\pm$ and $b$ would be hidden
in the
short-distance Wilson coefficients of local operators built out of the
$u,d,s$ and
$c$ quarks only.

The contributions from the tree-level ($T$),
annihilation
$(A)$, penguins $(\Delta P)$\footnote{If we neglect the (multi-)
Cabibbo-suppressed $b$ quark contribution, there are two diagrams to
consider with
opposite signs: the ``chin" of the penguin is either a
$d$ quark or a
$s$ quark. In the limit where $m_d = m_s, \Delta P \equiv P_s - P_d = 0$},
exchanges with
either a
$u\bar u$ or a
$d\bar d$ pair created $(E)$ and, finally, exchanges with a $s\bar s$ pair
created
$(E_s)$ lead to the relations
\ba
w_0       &=& T + \Delta P + 2E - E_s
\\
w_1 &=& T + \Delta P + \frac{1}{3} E_s - \frac{2}{3} A
\\
v_1 &=&  \frac{2}{3} E_s + \frac{2}{3} A.
\ea

If one assumes
\be
E = E_s
\ee
which is what one expects in the $SU(3)$ limit, then Eqs (35)-(37) imply
\be
w_0 = w_1 + v_1
\ee
and Eqs (32)-(34) now read
\ba
A (D^+ \to K^+ \bar K^0) &=& w_0 - \frac{3v_1}{2} \\
A (D^\circ \to K^+ K^-) &=& \frac{w_0}{2} (1+e^{-i\delta_{K\bar K}})
\\
A (D^\circ \to K^\circ \bar K^\circ) &=& - \frac{w_0}{2} (1-
e^{-i\delta_{K\bar K}}).
\ea
Eqs (41)-(42) are in good agreement with experimental data when Eq.(24) is used.

It is difficult to imagine a more spectacular illustration of  final state
hadronic effects
\cite{10} than Eqs (41)-(42). Furthermore, from the experimental value
\be
\frac{\Gamma (D^0 \to K^+ K^-)}{\Gamma (D^+ \to K^+ \bar K^0)} \simeq 1.6
\ee
one deduces $v_1 \simeq \frac{1}{6} w_0$ or, in terms of quark diagrams,
\be
\frac{E+A}{T+\Delta P + E} \simeq \frac{1}{4}.
\ee
Since $A$ and $\Delta P$ are expected to be quite small in our phenomenological
picture, Eq.(44) entails
\be
\frac{E}{T} \simeq \frac{1}{3}.
\ee
This result may be at odds with some theoretical prejudices but is required
by the
data : with $\delta_{K\bar K}$ given by Eq.(24), all the $D \to K \bar K$
data are
indeed very nicely fitted by Eqs (40)-(42) provided Eq.(44) holds.

The detailed analysis of $D \to K \bar K$ decays presented in this section
can be
repeated for $(D \to \pi\pi)$ and $(D \to K \pi)$ channels. In these channels,
sizeable color-suppressed quark diagrams (C) operate and nothing as
striking as Eq.(42) or as
unexpected as Eq.(45) emerges from such an analysis.

To summarize our earlier work on $(K\pi)$ channels as well as the results of the
present paper on $(\pi\pi)$ and $(K \bar K)$ decays let us insist on the
following
points:

- at the $D$ mass, hadronic phases are rather well estimated in the context of a
Regge model. Note that the hierarchy
\be
\delta_{K\pi} \simeq \frac{\pi}{2},  \ \
\delta_{\pi\pi} \simeq \frac{\pi}{3}, \ \
\delta_{K\bar K} \simeq -\frac{\pi}{6}
\ee
follows
from the difference in $u$-channel exchanges for the corresponding scattering
processes combined with different Clebsch Gordan coefficients weighing the
relative
contributions of the Pomeron and the $I=1$ Regge trajectories. In this
paper, we have ignored
inelastic channels such as $\{K\eta\} \to \{K \pi\}$ or $\{\pi \eta\} \to
\{K \bar K\}$. In
fact,  our Regge analysis shows that they have little effect on phases at
least at the $D$
mass.

- at the $B$ mass, hadronic phases are predicted to be non negligible in
the three
channels considered so far:
\be
\delta_{K\pi} \approx 17^\circ, \ \
\delta_{\pi\pi} \simeq 11^\circ, \ \
\delta_{K\overline K}
\approx -7^\circ.
\ee
We have assumed that inelastic channels have a small overall effect on
these hadronic phases
\cite{11}. Whether this is true or not is an experimental question. But
clearly final state
hadronic phases remain large in $B$ decays and the prospect for
CP-asymmetries looks
particularly promising in the $K \pi$ channel.

- the parametrization suggested  in Ref \cite{8} works very nicely as
exemplified by
our analysis of $(D \to K \bar K)$ decays. When hadronic phases are important,
quark-diagram absorptive parts and inelastic effects on the phases seem to be
negligible. These latter conclusions may not hold for decay
processes where hadronic phases are quite small.

\section*{Acknowledgements}

We have benefited from discussions with D. Del\'epine and C. Smith.

\end{document}